# High Performance Single Layered WSe$_2$ *p*-FETs with Chemically Doped Contacts

*Hui Fang[1,2,3], Steven Chuang[1,2,3], Ting Chia Chang[1], Kuniharu Takei[1,2,3], Toshitake Takahashi[1,2,3], and Ali Javey[1,2,3,\*]*

[1]*Electrical Engineering and Computer Sciences, University of California, Berkeley, CA, 94720.*

[2]*Materials Sciences Division, Lawrence Berkeley National Laboratory, Berkeley, CA 94720.*

[3]*Berkeley Sensor and Actuator Center, University of California, Berkeley, CA, 94720.*

\* *Correspondence should be addressed to A.J. ([ajavey@eecs.berkeley.edu](ajavey@eecs.berkeley.edu)).*

**Abstract** - We report high performance *p*-type field-effect transistors based on single layered (thickness, ~0.7 nm) WSe$_2$ as the active channel with chemically doped source/drain contacts and high-κ gate dielectrics. The top-gated monolayer transistors exhibit a high effective hole mobility of ~250 cm$^2$/Vs, perfect subthreshold swing of ~60 mV/dec, and $I_{ON}/I_{OFF}$ of >10$^6$ at room temperature. Special attention is given to lowering the contact resistance for hole injection by using high work function Pd contacts along with degenerate surface doping of the contacts by patterned NO$_2$ chemisorption on WSe$_2$. The results here present a promising material system and device architecture for *p*-type monolayer transistors with excellent characteristics.

Exploratory research is needed to develop materials, structures, and device technologies for future sub-5 nm gate length field-effect transistors (FETs). At such small scales, severe short channel effects limit the performance and operation of electronic devices[1]. Theoretical studies have shown that the use of large band gap semiconductors with ultrathin bodies and/or gate-all around structures are essential to minimize the short channel effects at extreme scaling limits[1]. Specifically, for ultrathin body devices, a general guideline dictates a body thickness of < 1/3 of the gate length for effective electrostatic control of the channel by the gate electrode[3]. For sub-5 nm gate lengths, this corresponds to channel materials with only 1-2 atomic layers in thickness. In this regard, single layered semiconductors are excellent candidates for the channel material of future monolayer-FETs (ML-FETs). As compared to materials with diamond/zinc-blende structure, layered semiconductors exhibit advantageous surfaces with minimal roughness, dangling bonds, defect states and native oxides. Among various layered materials, graphene has achieved worldwide interest and numerous interesting applications have been proposed[4,5,6,7]. However, graphene does not have an intrinsic band gap, which severely limits its digital logic applications. Meanwhile, the band gap of graphene nanoribbons shown to date is still too small for ultrashort channel FETs[8]. Recent advances in monolayer and few layered $MoS_2$, a chalcogenide semiconductor with a large band gap of $E_g$~1.8 eV, has shown the potential use of layered semiconductors for high performance *n*-type FETs (*n*-FETs)[9]. To date, however, high mobility monolayer *p*-FETs with high $I_{ON}/I_{OFF}$ have not been reported, and more importantly routes for controllable doping of chalcogenide layered semiconductors at the source/drain (S/D) contacts for low parasitic resistances have not been explored. In this report, by developing a surface dopant-profiling technique, we demonstrate the first high hole mobility $WSe_2$ ML-FETs (body thickness of ~0.7 nm) with degenerately doped contacts. The use of heavily p-doped contacts is essential in lowering the metal contact resistances to $WSe_2$ by orders of magnitude, and enabling the demonstration of *p*-FETs with peak effective mobility of ~250 cm$^2$/Vs, near ideal subthreshold swing (SS) of ~60 mV/decade, and high $I_{ON}/I_{OFF}$ of >10$^6$.

$WSe_2$ is a layered semiconductor with a bulk indirect bandgap of ~1.2 eV (ref. 10-11). A recent study of bulk $WSe_2$ FETs revealed an intrinsic hole mobility of up to 500 cm$^2$/Vs (ref. 12). However, the bulk devices exhibited poor $I_{ON}/I_{OFF}$ ratio of less than 10 at room temperature, along with ambipolar behavior, both of which are highly undesirable for digital logic applications. This is presumably due to the use of a bulk (i.e., thick) body which results in large OFF state leakage currents. Here we applied the well-known mechanical exfoliation method to obtain a single layer of $WSe_2$ from a bulk crystal (Nanoscience Instruments, Inc) on Si/SiO$_2$ substrates for ML-FET fabrication and characterization. Fig. 1a shows an optical microscope image of a single layer $WSe_2$ flake (light orange) transferred on top of a Si/SiO$_2$ (thickness, 270 nm) substrate. This thickness of SiO$_2$ is optimized to optically visualize the contrast of single layer and few layer $WSe_2$, similar to the cases of graphene and $MoS_2$ (ref. 9). Fig. 1b depicts the atomic force microscope (AFM) image of a single layered $WSe_2$ flake, with Fig. 1c showing the lateral height profile at the edge of the flake. From AFM measurements, the thickness of the single layer is determined to be ~0.7 nm, which agrees with the crystallography data of $WSe_2$ in literature[13]. Note that the surface roughness of $WSe_2$ is similar to the Si/SiO$_2$ background, indicating that the layer is uniform and the surface roughness is minimal, which is essential for obtaining high carrier mobilities with low surface roughness scattering rates.

Large $E_g$ semiconductors such as $WSe_2$ are notoriously known for their difficulty in forming ohmic metal contacts. Therefore, it is important to shed light on the metal-$WSe_2$ Schottky barriers (SBs), and explore routes towards minimizing the contact resistances to enable exploration of intrinsic material and device properties. In this regard, we have explored different metal source/drain (S/D) contacts, including Pd, Ag, Ni, Au, Ti and Gd for back-gated $WSe_2$ FETs. The fabrication process involves the transfer of $WSe_2$ layers onto a $Si/SiO_2$ substrate, followed by a 1hr acetone bath to remove the tape residues, S/D metal contact patterning by lithography, evaporation and lift off processes. Here the S/D length is fixed at $L$~8 µm. Based on the various metal contacts explored, high work function Pd was found to form the lowest resistance contact to the valence band of $WSe_2$ for hole transport, with devices exhibiting the highest unit-width normalized ON currents. As depicted in the back-gated transfer characteristics (Fig. 2a), Pd contacted FETs exhibit clear p-type conduction, without ambipolar transport. In contrast, lower work function metal contacts resulted in FETs that conduct in both n and p regimes with low current levels, reflecting high SB heights to both conduction and valence bands of $WSe_2$. Specifically, Ti forms near mid-gap SBs to $WSe_2$ with low-current ambipolar characteristics (Fig. 2a-b). The results here highlight the importance of selecting high work function metals with good interfaces to $WSe_2$ in order to lower the SB height at the contacts for hole transport. Clearly, Fermi level pinning is weak or non-existent at metal-$WSe_2$ interfaces. Further investigation of the exact effect of the metal work function and interface chemistry on the barrier height is needed in the future.

Although Pd was found to form the best contact for hole transport among the various metals explored, a small SB may still exist at the Pd-$WSe_2$ interface given the large $E_g$ of $WSe_2$. To shed light on the contact properties of Pd, surface hole doping of $WSe_2$ was explored. By heavily *p*-doping $WSe_2$, the width of any barriers at the metal interfaces can be drastically reduced, resulting in more efficient tunneling of the carriers and lower resistance contacts. Inspired by the surface doping approach in carbon nanotubes and graphene[14,15,16], here we utilized $NO_2$ molecules as a *p*-type surface dopant. $NO_2$ molecules are expected to be absorbed both physically and chemically on top of the $WSe_2$ surface as illustrated in Fig. 3a. Due to the strong oxidizing property of $NO_2$, the molecules act as "electron pumps" when chemisorbed to $WSe_2$. Fig. 3b shows the transfer characteristics of a back-gated $WSe_2$ ML-FET before and after $NO_2$ exposure. The device was exposed to 0.05% $NO_2$ in $N_2$ gas for 10 min, beyond which the doping effect was found to saturate presumably due to the $NO_2$ saturation coverage on the surface[14]. Here, the entire channel is exposed to $NO_2$, resulting in blanked (i.e., unpatterned) doping of $WSe_2$. The weak gate-voltage dependence of current after $NO_2$ exposure clearly reflects that $WSe_2$ is heavily doped (Fig. 3b). Moreover, the current at high negative $V_{GS}$ (ON state) is enhanced by >1000x after $NO_2$ doping, which can be attributed to the lowering of contact resistance by thinning the Pd-$WSe_2$ SB width for hole injection. In addition, $NO_2$ may increase the work function of Pd, thereby lowering the SB height at the interface. This work function increase is possibly due to the formation of surface/subsurface metastable palladium oxides when $NO_2$ is absorbed on Pd as previously reported in literature[17,18].

To estimate the 2-D sheet carrier density ($n_{2D}$) of $WSe_2$ after $NO_2$ doping, the source/drain current ($I_{DS}$) at zero gate voltage was modeled as $I_{DS} = q \cdot n_{2D} \cdot W \cdot \mu \cdot V_{DS} / L$, where q is the electron charge, *W* and *L* are the width and length of channel, respectively, $\mu$ is the field-effect mobility (~140 $cm^2$/Vs as extracted from the $I_{DS}$-$V_{GS}$ transfer characteristic), and $V_{DS}$ is the source/drain voltage. Since the channel shape is often irregular, the width is defined as the total channel area divided by the length. We note that the field-effect mobility from back-gated $WSe_2$ ML-FETs doped with $NO_2$ is 1-2 orders of magnitude higher than $MoS_2$ ML-FETs without the high-κ dielectric mobility booster[9]. This could be either due to the different surface characteristics of $WSe_2$ as compared to $MoS_2$ and/or due to the lower contact resistance observed here by doping the contacts. $n_{2D}$ is extracted to be ~$2.2\times10^{12}$ cm$^{-2}$, which corresponds to a doping concentration of ~$3.1\times10^{19}$ cm$^{-3}$. At this doping concentration, Fermi level lies at ~16 meV below the valence band edge ($E_V$), as calculated from the Joyce-Dixon Approximation[19] and an effective hole density of state of $N_V$=$2.54\times10^{19}$ cm$^{-3}$ (ref. 20). Therefore, NO$_2$ exposed WSe$_2$ layers are degenerately doped. This doping level, however, is lower than the NO$_2$ surface monolayer density of ~$1.4\times10^{15}$ cm$^{-2}$ (assuming a perfect monolayer coverage)[17], suggesting that on average ~0.001 electron is transferred per NO$_2$ molecule. It should be noted that NO$_2$ doping is reversible due to the gradual desorption of NO$_2$ molecules from the WSe$_2$ surface once exposed to ambient air (Fig. S1). In the future, other dopant species and/or process schemes should be explored for permanent doping.

Next, we explored patterned *p*-doping of WSe$_2$ for the fabrication of top-gated ML-FETs with self-aligned, chemically doped S/D contacts. Pd/Au (30/20 nm) metal contacts were first defined by lithography and metallization. Gate electrodes, underlapping the S/D by a distance of 300 - 500 nm were then patterned by e-beam lithography and using PMMA as resist, followed by atomic layer deposition (ALD, at 120ºC) of 17.5 nm ZrO$_2$ as the gate dielectric, the deposition of Pd metal gate, and finally lift-off of the entire gate stack in acetone. While it has been reported that direct ALD on pristine graphene is not possible due to the lack of dangling bonds, uniform ALD of Al$_2$O$_3$ and HfO$_2$ on MoS$_2$ at the optimized temperature window has been previously demonstrated and attributed to the physical absorption of precursors on the basal plane[21,22], which we assume also applies to WSe$_2$. The devices are then exposed to a NO$_2$ environment and measured. Fig. 4a shows the schematic illustration of a top-gated ML-FET after NO$_2$ S/D doping. The exposed (underlapped) regions are p-doped heavily, while the gated region remains near intrinsic due to the protection of the active channel by the gate stack. This p+/i/p+ device structure is similar to conventional ultra-thin body Si MOSFETs. Fig. 4b shows the transfer characteristics of a ~9.4 µm channel length WSe$_2$ ML-FET (see Fig. S2 for the device optical images) before and after NO$_2$ contact doping. Here the back-gate voltage is fixed at -40 V to electrostatically dope the underlapped regions for both before and after NO$_2$ exposure. As a result, the difference in the current-voltage characteristics for the two measurements purely arises from the change of the metal-WSe$_2$ contact resistance, rather than the resistance of the underlapped regions. As depicted in Fig. 4b, a drastic enhancement of ~1000x improvement in the ON current is observed in the device after surface doping of the contacts by NO$_2$, without a change in I$_{OFF}$. A small shift in the threshold voltage to the positive direction is observed after NO$_2$ contact doping, which could be due to the increase of the Pd metal gate work function by NO$_2$. The ML-FET with doped contacts exhibits an impressive $I_{ON}/I_{OFF}$ of >10$^6$ arising from the large band-gap of WSe$_2$ combined with the monolayer-thick body which minimizes OFF state leakage currents.

Importantly, the transfer characteristics at room temperature shows a perfect subthreshold swing (*SS*), reaching the theoretical limit of ln(10)× $kT/q$ =60 mV/dec for a MOSFET, which originates from the thermionic emission of the source holes with density of states (DOS) tailed by the Fermi-Dirac distribution. For an experimental (i.e., non-ideal) MOSFET, SS is given as η×60 mV/decade, where $\eta \approx 1 + \frac{C_{it}}{C_{ox}}$ is the body factor, and $C_{it}$ is the capacitance caused by the interface traps ( $C_{it} = D_{it} \cdot q^2$, with $D_{it}$ being the interface trap density) and $C_{ox} = \varepsilon_{ox}\varepsilon_0 / T_{ox}$ is the top gate oxide capacitance per unit area ($\varepsilon_{ox}$ ~12.5 is the dielectric constant of ZrO$_2$, $\varepsilon_0$ is the vacuum permittivity, and $T_{ox}$= 17.5 nm is the ZrO$_2$ thickness). The experimental SS~60 mV/decade for WSe$_2$ ML-FETs suggests the near unity η caused by $C_{it}$ <<$C_{ox}$. The low $C_{it}$ is attributed to the lack of surface dangling bonds for layered semiconductors. Notably our measured SS outperforms all Ge and III-V MOSFETs, firmly indicating that WSe$_2$ has optimal switching characteristics for low power and high speed electronics.

Next the effective hole mobility, $\mu_{eff}$, of top-gated WSe$_2$ ML-FETs with doped contacts was extracted from the I-V characteristics by using the relation, $\mu_{eff} = \frac{\partial I_{DS}}{\partial V_{DS}} \frac{L_G}{C_{ox}(V_{GS} - V_T - 0.5 V_{DS})}$, where $V_T$ is the threshold voltage and $L_G$ is the gate length. The long-channel device exhibits a peak hole effective mobility of ~250 cm$^2$/Vs (Fig. 4c). The mobility does not degrade severely at high fields (Fig. 4c), which should be attributed to the fact that the carriers are already close to the gate in a single layered channel, and that surface roughness is minimal. Therefore, the gate oxide thickness can be further scaled without severe mobility degradation, again indicating that WSe$_2$ is a promising candidate for future scaled electronics. It must be noted that the channel is one monolayer thick (~ 0.7 nm), significantly thinner than the previously reported high hole mobility III-V or Ge MOSFETs, even in ultrathin body (UTB) configuration. For conventional diamond/zinc-blende structured material channels, severe mobility degradation occurs when reducing the channel thickness due to the enhanced scattering from both surface roughness and dangling bonds. For example, the peak effective hole mobility of strained-InGaSb based UTB-FETs drops from ~820 to 480 cm$^2$/Vs when the body thickness is reduced from 15 nm to 7 nm[23]. Therefore, the measured mobility of 250 cm$^2$/Vs for our monolayer-thick WSe$_2$ FETs is impressive.

The output characteristic of the same top gated WSe$_2$ ML-FET with doped contacts is shown in Fig. 4d. The long-channel device exhibits clear current saturation at high V$_{DS}$ due to pinch-off, similar to the conventional MOSFETs. In the low V$_{DS}$ regime, the I-V curves are linear, depicting the ohmic metal contacts. Overall, the results here demonstrate the potential of WSe$_2$ monolayers along with the essential patterned doping of the contacts for high performance *p*-FETs.

In conclusion, the layered semiconductor WSe$_2$, has been thinned down to a single layer through mechanical exfoliation and fabricated into *p*-FETs with promising hole mobility and perfect subthreshold characteristics. A NO$_2$ surface doping strategy is introduced to degenerately dope the S/D regions of the FETs and drastically reduce the metal contact resistance, meanwhile revealing intrinsic transport properties of the channel. Along with the previously demonstrated MoS$_2$ single layer transistor, the results encourage further investigation of layered semiconductors, especially the transition metal dichalcogenide family, for future high performance electronics. As emphasized in this work, surface doping is a necessity for obtaining high performance ML-FETs, and in this regard exploration of other dopant species for both n- and p-doping in needed in the future.

## Acknowledgements

This work was funded by NSF E3S Center and FCRP/MSD. The materials characterization part of this work was partially supported by the Director, Office of Science, Office of Basic Energy Sciences, and Division of Materials Sciences and Engineering of the U.S. Department of Energy under Contract No. De-Ac02-05Ch11231 and the Electronic Materials (E-Mat) program. A.J. acknowledges a Sloan Research Fellowship, NSF CAREER Award, and support from the World Class University program at Sunchon National University.

**SUPPORTING INFORMATION**

Reversibility of $NO_2$ doping; optical microscope images of a top-gated ML-FET. This material is available free of charge via the Internet at http://pubs.acs.org.

**Figure Captions**

**Figure 1. Single layered WSe$_2$ on a Si/SiO$_2$ substrate. a**, Optical microscope image of a single layered WSe$_2$ (light orange flake) on a Si substrate with 270 nm SiO$_2$. **b**, AFM image of a single layered WSe$_2$ on Si/SiO$_2$. **c**, Height profile of a line scan (as indicated by the dashed line in Fig. 1b) across the single layered WSe$_2$-SiO$_2$ boundary.

**Figure 2. Back-gated WSe$_2$ FETs with different metal contacts. a**, $I_{DS}$-$V_{GS}$ characteristics of Pd (red curve) and Ti (black curve) contacted WSe$_2$ FETs on a Si substrate with 50 nm SiO$_2$ as the back-gate dielectric. Here WSe$_2$ is few layered (thickness, ~5 nm). **b**, Qualitative energy band diagrams for Pd (top) and Ti (bottom) contacted WSe$_2$ FETs in the ON-state, depicting the height of the SBs for hole injection ($\Phi_{Bp}$) at the metal-WSe$_2$ interfaces.

**Figure 3. Chemical *p*-doping of single layered WSe$_2$ by NO$_2$. a**, Cross-sectional schematic of a back-gated WSe$_2$ ML-FET on Si/SiO$_2$, with NO$_2$ molecules being absorbed on both the channel and contacts. **b**, $I_{DS}$-$V_{GS}$ characteristics of Pd contacted WSe$_2$ ML-FET before (black curve) and after (red curve) exposure to NO$_2$.

**Figure 4. Top-gated WSe$_2$ ML-FETs with chemically doped contacts. a**, Schematic of a top-gated WSe$_2$ ML-FET, with chemically *p*-doped S/D contacts by NO$_2$ exposure. Here the top-gate acts as the mask for protecting the active channel from NO$_2$ doping. **b**, Transfer characteristics of a device with *L*~9.4 µm, before and after NO$_2$ patterned doping of the S/D contacts. **c**, Extracted effective hole mobility as a function of gate overdrive of the device shown in b at $V_{DS}$= -0.05 V. **d**, Output characteristics of the same device shown in b.

# References

1. Taur, Y. *IBM J. Rev & Dev.* **2002**, 46, 213-222.

2. Luisier, M.; Lundstrom, M.; Antoniadis, D. A.; Bokor, J. *IEDM Tech. Dig.* **2011**, 251-254.

3. Chau, R.; Kavalieros, J.; Doyle, B.; Paulsen, N.; Lionberger, D.; Barlage, D.; Arghavani, R.; Roberds, B.; Doczy, M. *IEDM Tech. Dig.* **2001**, 621-624.

4. Liu, M.; Yin, X.; Ulin-Avila, E.; Geng, B.; Zentgraf, T.; Ju, L.; Wang, F.; Zhang, X. *Nature* **2011**, 474, 64-67.

5. Schedin, F.; Geim, a K.; Morozov, S. V.; Hill, E. W.; Blake, P.; Katsnelson, M. I.; Novoselov, K. S. *Nat. Mater.* **2007**, 6, 652-655.

6. Frank, O.; Tsoukleri, G.; Riaz, I.; Papagelis, K.; Parthenios, J.; Ferrari, A. C.; Geim, A. K.; Novoselov, K. S.; Galiotis, C. *Nat. Commun.* **2011**, 2, 255.

7. Britnell, L.; Gorbachev, R. V.; Jalil, R.; Belle, B. D.; Schedin, F.; Mishchenko, A.; Georgiou, T.; Katsnelson, M. I.; Eaves, L.; Morozov, S. V.; Peres, N. M. R.; Leist, J.; Geim, A. K.; Novoselov, K. S.; Ponomarenko, L. A. *Science* **2012**, 335, 947-950.

8. Wang, X.; Ouyang, Y.; Jiao, L.; Wang, H.; Xie, L.; Wu, J.; Guo, J.; Dai, H. *Nat. Nanotech.* **2011**, 6, 563-567.

9. Radisavljevic, B.; Radenovic, A; Brivio, J.; Giacometti, V.; Kis, A. *Nat. Nanotech.* **2011**, 6, 147-150.

10. Upadhyayula, L. C.; Lorerski, J. J.; Wold, A.; Giriat, W.; Kershaw, R. *J. Appl. Phys.* **1968**, 39, 4736-4740.

11. Yousefi, G. H. *Mater. Lett.* **1989**, 9, 38-40.

12. Podzorov, V.; Gershenson, M. E.; Kloc, C.; Zeis, R.; Bucher, E. *Appl. Phys. Lett.* **2004**, 84, 3301-3303.

13. Kalikhman, V. L.; Umanskii, Y. S. *Sov. Phys. Usp.* **1973**, 15, 728-740.

14. Kong, J.; Franklin, N. R.; Zhou, C.; Chapline, M. G.; Peng, S.; Cho, K.; Dai, H. *Science* **2000**, 287, 622-625.

15. Chen, W.; Chen, S.; Qi, D. C.; Gao, X. Y.; Wee, A. T. S. *J. Am. Chem. Soc.* **2007**, 129, 10418-10422.

16. Wehling, T. O.; Novoselov, K. S.; Morozov, S. V.; Vdovin, E. E.; Katsnelson, M. I.; Geim, A. K.; Lichtenstein, A. I. *Nano Lett.* **2008**, 8, 173-177.

17. Bartram, M. E.; Windham, R. G.; Koel, B. E. *Surf. Sci.* **1987**, 184, 57-74.

18. He, J.-W.; Memmert, U.; Norton, P. R. *J. Chem. Phys.* **1989**, 90, 5088-5093.

19. Joyce, W. B.; Dixon, R. W. *Appl. Phys. Lett.* **1977**, 31, 354-356.

20. Spah, R.; Lux-steiner, M.; Obergfell, M.; Ucher, E.; Wagner, S. *Appl. Phys. Lett.* **1985**, 47, 871-873.

21. Liu, H.; Ye, P. D. *IEEE Electron Device Lett.* **2012**, 33, 546-548.

22. Liu, H.; Xu, K.; Zhang, X.; Ye, P. D. *Appl. Phys. Lett.* **2012**, 100, 152115.

23. Takei, K.; Madsen, M.; Fang, H.; Kapadia, R.; Chuang, S.; Kim, H. S.; Liu, C.-H.; Plis, E.; Nah, J.; Krishna, S.; Chueh, Y.-L.; Guo, J.; Javey, A. *Nano Lett.* **2012**, 12, 2060-2066.

**Figure 1**

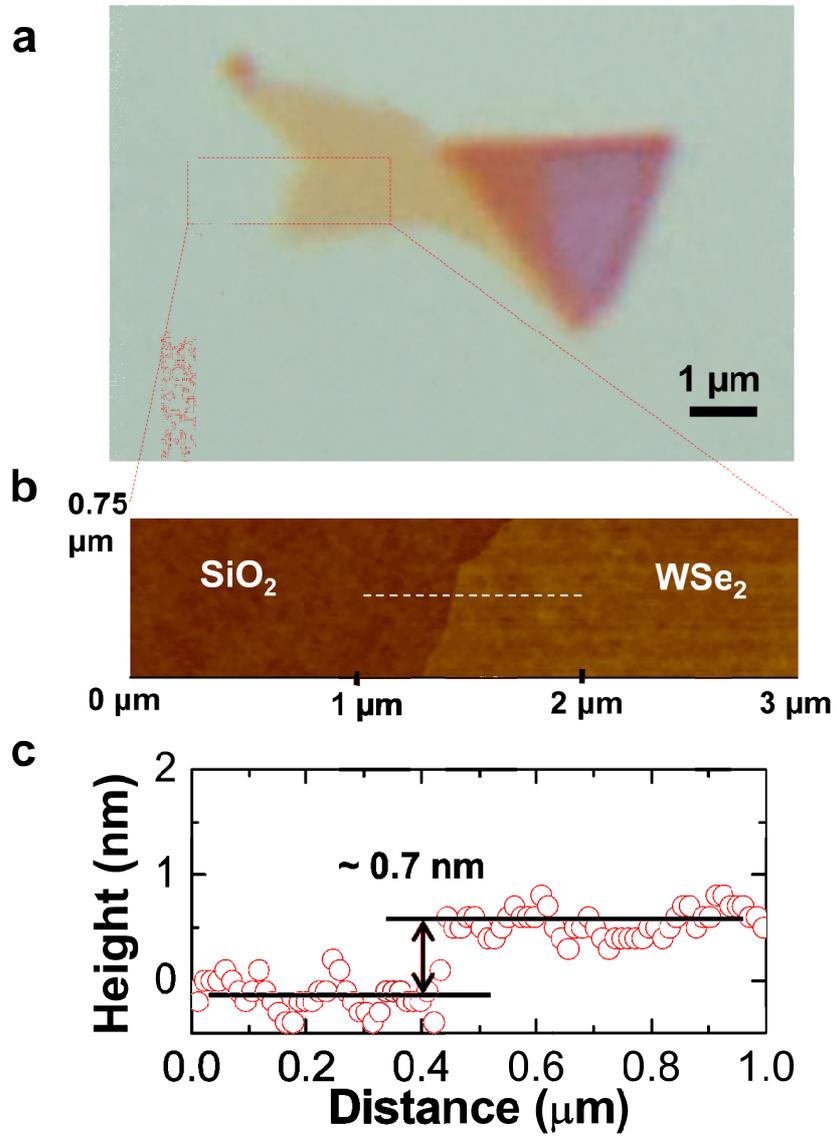

**Figure 2**

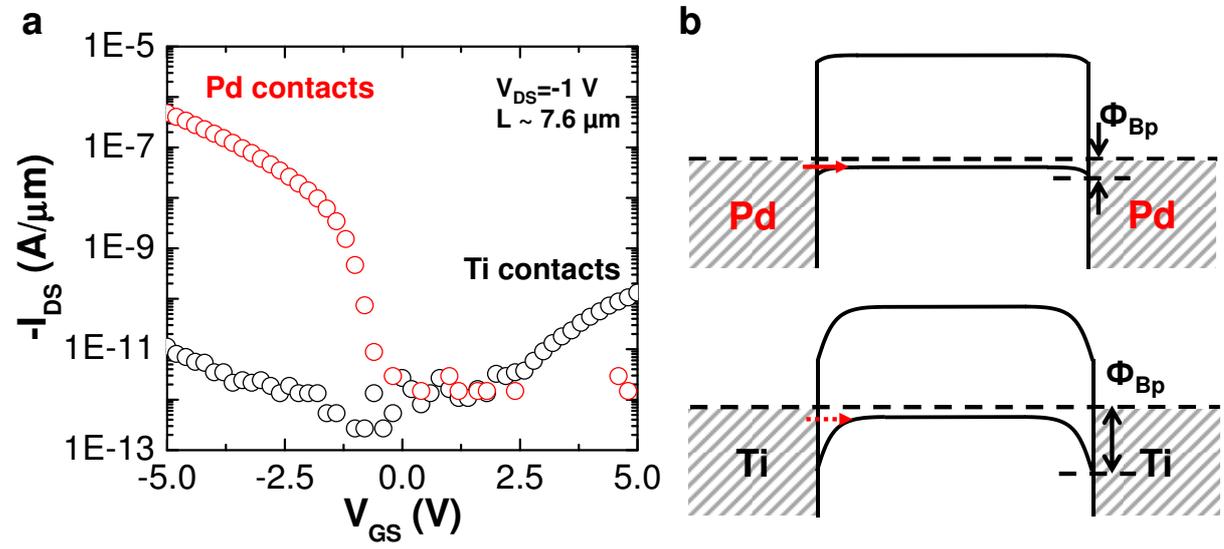

**Figure 3**

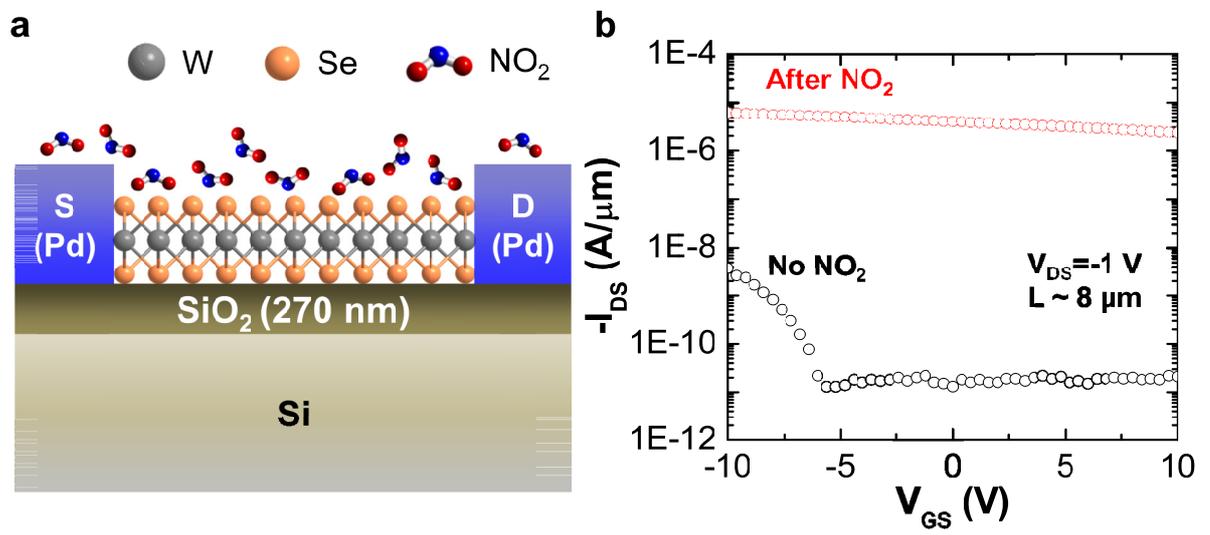

Figure 4

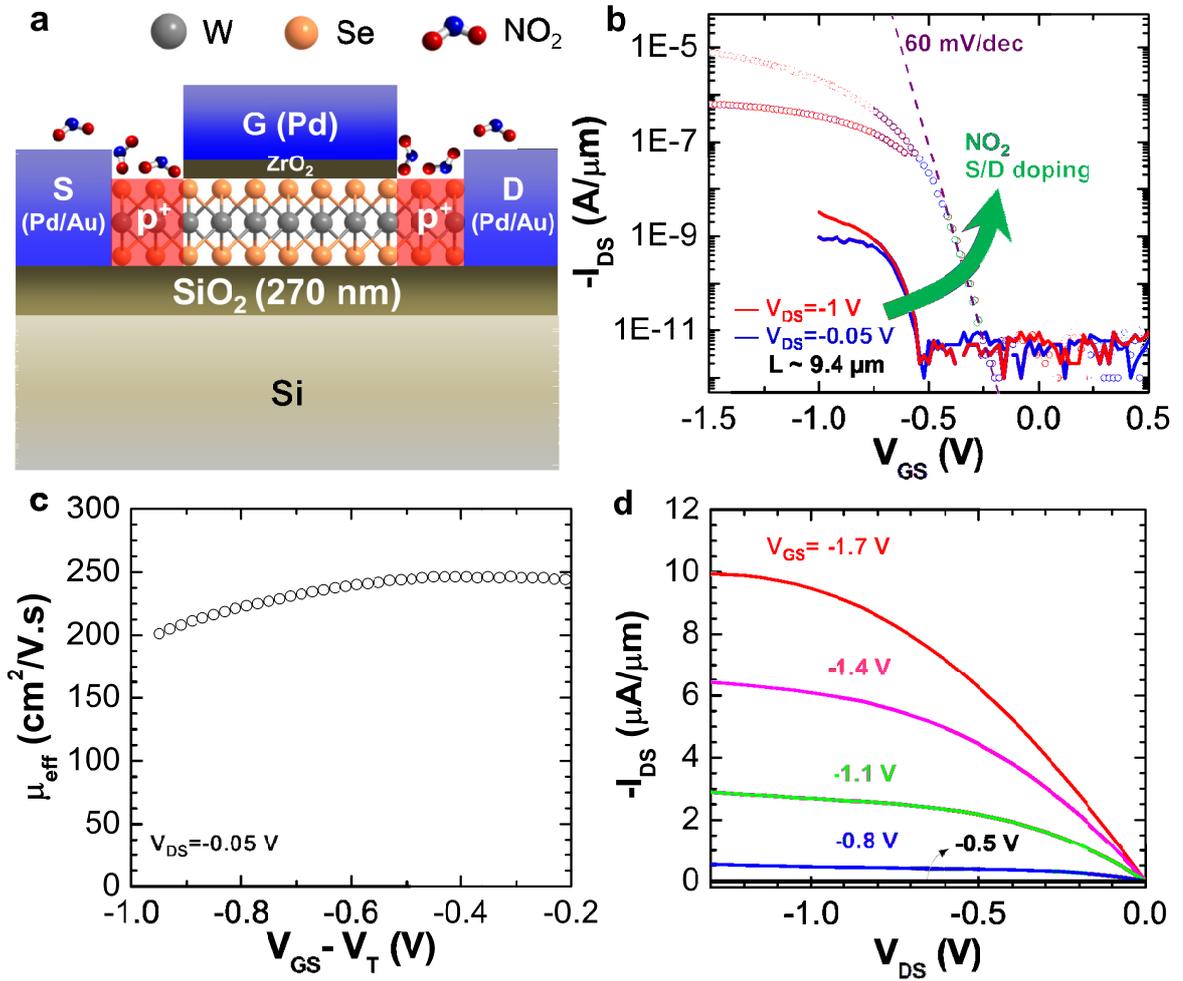